\documentclass[prl,preprint,showpacs,groupedaddress]{revtex4}
\usepackage{dcolumn}
\usepackage{graphics}
\usepackage{amsmath}
\usepackage{bm}

\makeatletter
\def\btt#1{\texttt{\@backslashchar#1}}%
\DeclareRobustCommand\bblash{\btt{\@backslashchar}}%
\makeatother

\begin{document}


\title[Short Title]{The effect of a nucleating agent on lamellar growth \\
in melt-crystallizing polyethylene oxide}

\author{F. Aliotta $^1$}
\author{G. Di Marco $^1$}
\author{R. Ober $^2$}
\author{M. Pieruccini $^1$}
\affiliation{%
$^1$ C.N.R. Istituto per i Processi Chimico-Fisici, sez. Messina\\
via La Farina 237, I-98123 Messina, Italy
}%
\affiliation{%
$^2$ Laboratoire de Physique de la Matiere Condensee, C.N.R.S. URA 792,
College de France;\\
11, Place Marcelin Berthelot,
F-75231 Paris Cedex, France
}%

\begin{abstract}
The effects of a (non co-crystallizing) nucleating agent on secondary nucleation rate and final lamellar thickness in isothermally melt-crystallizing polyethylene oxide are considered. SAXS reveals that lamellae formed in nucleated samples are thinner than in the pure samples crystallized at the same undercoolings. These results are in quantitative agreement with growth rate data obtained by calorimetry, and are interpreted as the effect of a local decrease of the basal surface tension, determined mainly by the nucleant molecules diffused out of the regions being about to crystallize. Quantitative agreement with a simple lattice model allows for some interpretation of the mechanism.
\end{abstract}


\maketitle

Nucleating agents have important applications in the industrial use of polymeric materials. As an example, mechanical properties such as hardness and elastic moduli can be controlled with these additives, which are able to affect crystallization kinetics rather dramatically \cite{1}. In fact, not only the \textit{primary} nucleation rate is influenced, but also the lamellar growth process via \textit{secondary} nucleation. In the context of bulk crystallization, the latter mechanism has been envisaged for the first time within the framework of a phenomenological interpretation of the free growth kinetics' data of indigo-nucleated isotactic polypropylene \cite{2,3}. Later, similar observations in nucleated high density polyethylene and polyethylene oxide (PEO) have been interpreted in the same way, with the further support of lattice calculations for the basal lamellar interface \cite{4}.

The basic idea is that during growth, the nucleant molecules present in regions which are about to crystallize, diffuse towards the forming basal interfaces prior to crystallization (in all the cases above, the nucleating agent did not co-crystallize). These molecules cause a local decrease of the basal surface tension $\sigma_e$, so they tend to remain confined in the interfacial regions because diffusion towards the amorphous bulk would require an extra energy to restore a higher value of $\sigma_e$. In this situation the linear growth velocity $v$ in low to moderate undercooled melts, is still related to the secondary nucleation free enthalpy $\Delta G$ by \cite{5}
\begin{equation}
	v\propto \exp \{-\Delta G/kT\}
\end{equation}
(with $T$ the temperature), but the expression
\begin{equation}
	\sigma_e=\sigma_{e0}\left(1+\frac{a}{\Delta T}\right),
\end{equation}
has to be substituted for $\sigma_{e0}$ in the formula
\begin{equation}
	\Delta G=4 b_0 \sigma \sigma_{e0} \frac{v_c T_m}{H_f \Delta T}
\end{equation}
customarily used for pure systems. In the above Eqs. 2 and 3, $b_0$ is the stem diameter, $\sigma$ is the lateral surface tension, $\sigma_{e0}$ is the basal surface tension in the absence of the nucleant, $T_m$ is the equilibrium melting point, $H_f/v_c$ is the enthalpy of fusion per unit volume, $\Delta T =T_m-T$ and $a$ ($<0$) is a coefficient proportional to the concentration of nucleating agent \cite{2,3,4}.

 \begin{figure} \begin{center}\scalebox{0.8}
{\includegraphics{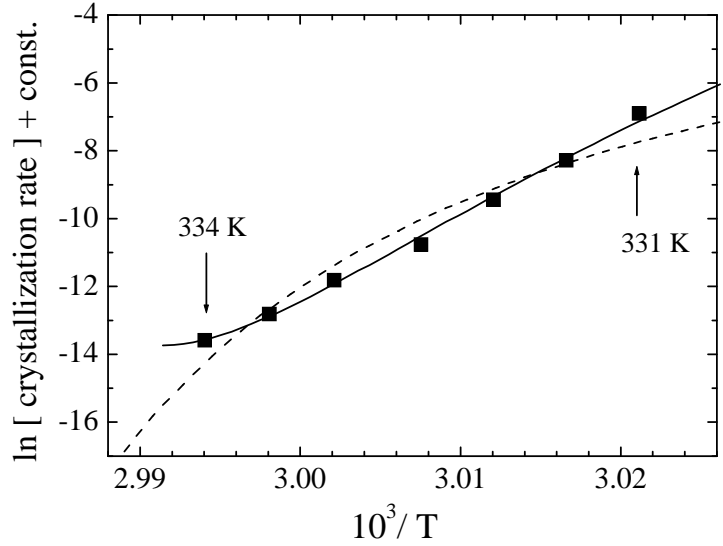}}
 \caption{
Logarithm of the crystallization rate (approximately $\ln v^2$) as a function of $T^{-1}$ for PEO nucleated with a number density $n=2.6\times10^{19}$ cm$^{-3}$ of indigo; the dashed line is a best fit using Eqs. 1 and 3; the solid line is obtained after substitution of $\sigma_{e0}$ with Eq. 2.
 }
 \label{fig1}\end{center}
 \end{figure}

Fig.1 shows the overall crystallization rates obtained by accurate DSC measurements on PEO nucleated with a number density $n\cong 2.6\times 10^{19}$ cm$^{-3}$ of indigo, i.e. 1\% wt fraction, in the interval $331\leq T \leq 334$ K  \cite{4}. Fittings with Eqs. 1-3 are also reported [i.e. assuming a constant nucleation density $N(\Delta T)$ in this small range] for the cases where either $a$ is let free to change, or is set equal zero. With $a\cong -1.8$ K ($T_m = 338$ K) the fitting improves significantly \cite{4}; a close value is also obtained with a simple lattice model \cite{4}. All this, and the fact that $\sigma_{e0}-\sigma_e$ \textit{decreases} with $\Delta T$, implies that the mechanism described by Eq. 2 overwhelmes the effect of an \textit{increasing} $N(\Delta T)$ in this range (the irrelevance of other possibly competing mechanisms has been briefly discussed in \cite{4}).

Consider now the morphological features controlled by the nucleant in \textit{crystallized} PEO. At fixed $\Delta T$, number and final size of the spherulites are influenced by indigo through the primary nucleation rate. About a morphological counterpart of the secondary nucleation route to crystallization kinetics' control, a fundamental role is played by Eq. 2. Since the lamellar fold length is related to $\sigma_e$ by
\begin{equation}
	l = 2\sigma_{e}\frac{v_c T_m}{H_f \Delta T},
\end{equation}
it is expected that the lamellar thickness of nucleated samples should be smaller than in pure samples crystallized at the same $\Delta T$, and for $\left|a/\Delta T\right|$ not too large:
\begin{equation}
	\frac{l_0 - l}{l_0}\cong - \frac{a}{\Delta T},
\end{equation}
where $l_0$ is given by Eq. 4 with $a=0$.

SAXS measurements carried out on pure and nucleated PEO (with the same content of indigo as in Fig. 1) in fact confirm this prediction.

To assure homogeneity, nucleated samples were prepared from a solution of PEO 600 000 (Aldrich) and the appropriate amount of indigo in acetonitrile, which was kept under stirring for one day. The solvent was then very slowly evaporated at about 50 °C. Pure PEO samples were prepared with similar procedure. Before crystallization, the samples were kept above the melting temperature for half an hour or more, and then cooled as rapidly as possible to the preset crystallization temperature.

The SAXS measurement apparatus consists of a Rigaku rotating anode generator, with a wavelength of 1.54 $\mathrm \AA$. The counter was 76 cm far from the sample (held the latter at room temperature), and the dimension of the beam at the counter was $3\times0.5$ mm$^2$, so that desmearing was necessary (point collimation measurements, with a $0.5\times0.5$ mm$^2$ beam, were performed as a check, and gave the same results, although with less statistics).

The lamellar thicknesses in the pure ($l_0$) and nucleated ($l$) samples have been estimated from the evaluation of the correlation function $K(s)$ \cite{6}. The relatively small electron density difference between amorphous and crystalline regions in PEO rendered the data analysis a bit cumbersome. The spurious background forward scattering characterizing these systems (which is not unusual in general \cite{7}) was modelled by a \textit{local} (i.e. $q=0$) gaussian distribution of density fluctuations, whose amplitude and width (the latter always found in the range $4.5-7.5 \times 10^{-3} \mathrm\AA ^{-1}$) were obtained by means of best fitting procedures.
Fig. 2 shows the interface distributions $K"(s)$ for some of the PEO samples. The first maximum in $K"(s)$ is assigned to the interlamellar thickness, since \textit{i)} its position is weakly dependent on $\Delta T$, as can be seen from Tab. 1, and \textit{ii)} the linear crystallinity thus found gets closer to the DSC one ($\approx$ 65 \% at a rate -1 °C/min in both nucleated and pure PEO). The resulting Gibbs-Thompson plot is in Fig. 3. The linear fit to the pure PEO data points extrapolates to a melting temperature $T_c^\infty=339.3$ K, i.e. rather close to the above value of $T_m$.

 \begin{figure} \begin{center}\scalebox{0.8}
{\includegraphics{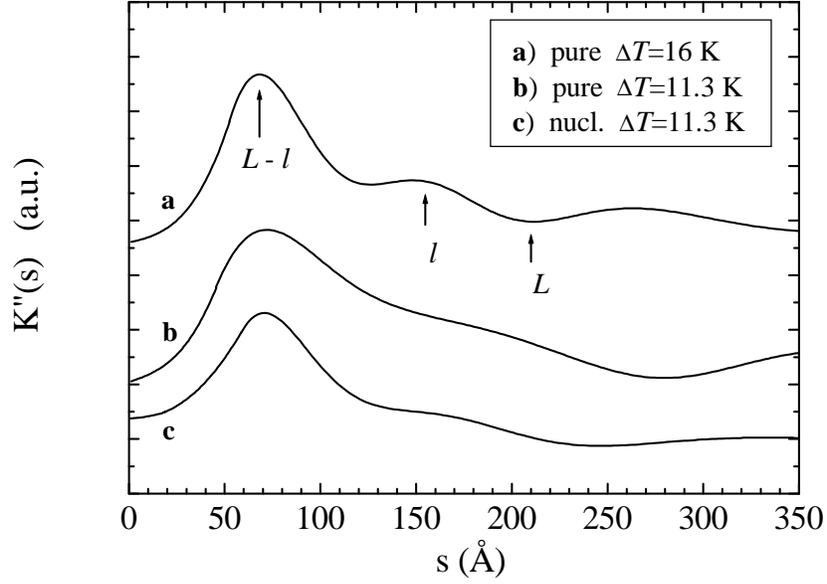}}
 \caption{
Comparison between interface distances distribution functions $K"(s)$ of pure and nucleated PEO samples crystallized at different $\Delta T$. $L$ and $l$ are the long period and the lamellar thickness respectively
 }
 \label{fig2}\end{center}
 \end{figure}

 \begin{figure} \begin{center}\scalebox{0.8}
{\includegraphics{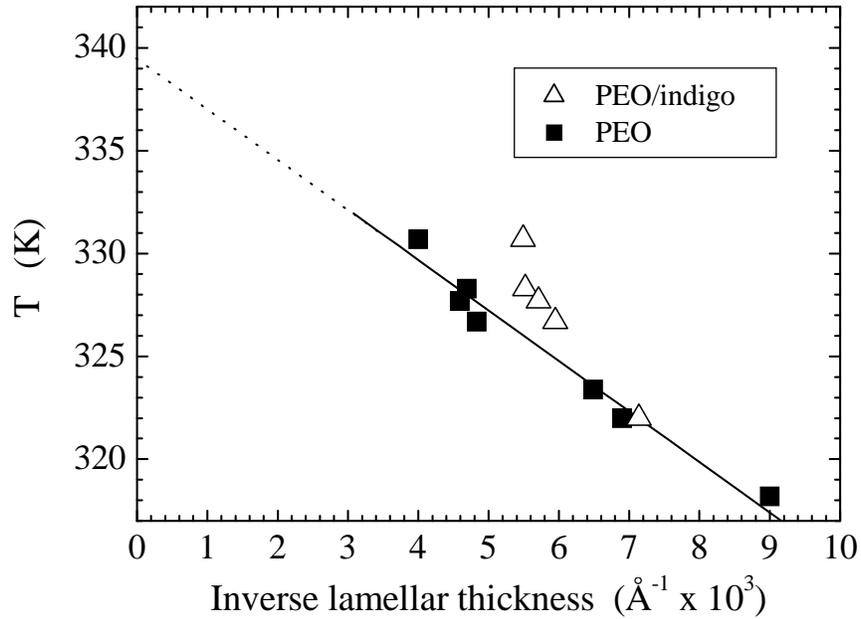}}
 \caption{
Inverse lamellar thickness as a function of the temperature for pure and nucleated PEO from SAXS.
}
 \label{fig3}\end{center}
 \end{figure}

As the crystallization temperature decreased, complete thermalization was ever more difficult to achieve in PEO/indigo samples before the onset of crystallization. For nucleated samples crystallized at large $\Delta T$ the preset temperature was only apparent. For example, at $T=322$ K the lamellar thickness of the pure and nucleated samples coincide, although a 0.1 relative difference had to be expected.

The quantity $-a/\Delta T$ as obtained from SAXS, fitting of Fig. 1 data (Cal.) and from lattice calculations, is reported in Tab. 1. The good agreement between the values of this parameter as obtained by SAXS (\textit{grown} lamellae) and best fitting (\textit{growing} lamellae), indicates that in the $T$-range considered the mechanism described by Eq. 2 plays a main role in the influence of the nucleant on the growth process (in particular, effects of nucleation density changes are negligible).
Moreover, it supports the hypothesis that the nucleant remains confined at the basal interfaces during lamellar growth, without appreciable increase of the nucleant concentration in the remaining amorphous phase. 

\begin{table}
\caption{SAXS results for long period and lamellar thickness in pure ($L_0$ and $l_0$) and nucleated ($L$ and $l$) PEO, all in \AA\ , at the corresponding undercoolings
$\Delta T$. The estimates
of $-a/\Delta T$ are obtained from SAXS, best fit in Fig. 1 (Cal.) and lattice
calculations with $\sigma_{e0}=20$ erg cm$^{-2}$ (Lat$_1$) and $\sigma_{e0}=70$ erg cm$^{-2}$ (Lat$_2$).}
\begin{center}
\renewcommand{\arraystretch}{1.4}
\begin{ruledtabular}
\begin{tabular}{cccccccccccccc}
 	 $\Delta T$ &
   &
   &
   $L_0$ &
   $l_0$ &
   &
   $L$ &
   $l$ &
   &
   &
  $-a/\Delta T$ &
   &
   \\
\noalign{\smallskip} \hline \noalign{\smallskip}

   &
   &
   &
   &
   &
   &
   &
   &
   &
   SAXS &
   Cal. &
   Lat$_1$ &
   Lat$_2$ \\
\noalign{\smallskip} \hline \noalign{\smallskip}
 7.3 & & & 330 & 250 & & 260 & 180 & & 0.28 & 0.25 & 0.24 & 0.17 \\
 9.7 & & & 285 & 215 & & 250 & 180 & & 0.16 & 0.18 & 0.17 & 0.12 \\
10.3 & & & 290 & 220 & & 240 & 175 & &  0.2 & 0.17 & 0.16 & 0.11 \\
11.3 & & & 280 & 210 & & 240 & 170 & & 0.14 & 0.16 & 0.15 & 0.11 \\
14.6 & & & 230 & 160 & &  -  &  -  & &   -  & 0.12 & 0.12 & 0.09 \\
  16 & & & 215 & 145 & & 210 & 140 & & 0.03 & 0.11 & 0.11 & 0.08\\
19.8 & & & 180 & 110 & &  -  &  -  & &   -  & 0.09 & 0.09 &  0.06\\
\end{tabular}
\end{ruledtabular}
\end{center}
\label{Tab1}
\end{table}

Some insight in the effect can be gained from the lattice description \cite{4}.
In the crystalline region the chain segment fractional occupation of the lattice sites is $\phi_c=1$, and the orientational entropy associated to bonds connecting adjacent segments in the chain is lowest. On the other hand, in the amorphous bulk the fractional occupation $\phi_a$ (i.e. the amorphous-to-crystalline density ratio) is less than $\phi_c$, and is related to an appropriate Flory-Huggins segment/vacuum interaction parameter $\chi_{p0}$ (=1.8 for PEO); correspondingly, the bond orientational entropy is highest. In the interfacial region the fractional occupation $\phi$ is higher than $\phi_a$, i.e. the chain segments are somewhat compressed, while a relatively large bond orientational entropy is still maintained. This contributes to the overall $\sigma_{e}$. Now, when a nucleant molecule is put in place of a segment in a site adjoining the folding plane, then \textit{i}) a nucleant/segment solvation energy arises, which is described by a parameter $\chi_{s}$ ($\leq 0$), \textit{ii}) a local segment-segment compression work correspondingly disappears and \textit{iii}) steric hindrance causes a decrease in the bond orientational entropy associated to the neighboring chain segments. These factors contribute to to the change in $\sigma_e$, which can be approximated by \cite{8}:
\begin{equation}
\left(\frac{d\sigma_e}{d\varphi}\right)_{n=0}\approx\frac{kT}{b_0^2}\left[\left(\chi_s-\chi_{p0}\right)-\frac{1}{2}-\ln \left(\frac{z}{2}\right)\right],
\end{equation}
where $z$ $(=6)$ is the lattice coordination number; $\varphi$ is the average number of nucleant molecules diffusing out of the volume of a forming stem, towards \textit{one} of the ends of the stem itself (for finite $\Delta T$ it is given by $\varphi \equiv n b_0^2 l_0 /2$); the value ~$-0.5$ can be taken for $\chi_s$, corresponding to a "solvation" energy of $\approx 1.3$ kJ/mol typical of this kind of "solutions".
Note that in this (rough) approximation $(\partial \sigma_e/\partial\varphi)_{n=0}$ is independent of $\sigma_{e0}$. In fact, the calculation of the \textit{full} expression doesn't show a dramatic dependence on $\sigma_{e0}$. As a practical example, assuming $\sigma_{e0}=20$ erg cm$^{-2}$ \cite{9} one finds $a\cong -1.7$ K from the expression
$a\cong n b_0^2 T_m v_c H_f^{-1}(\partial \sigma_e/\partial\varphi)_{n=0}$ \cite{4}; on the other hand, taking $H_f/v_c \cong 210$ J cm$^{-3}$ \cite{9, 10}, SAXS data imply $\sigma_{e0}\approx 70$ erg cm$^{-2}$ (cfr. \cite{10}), and the calculation yields $a\cong -1.2$ K in this case. In Table 1 the values of $-a/\Delta T$ for both these $\sigma_{e0}$ values are reported.

In summary, we have shown that in the present case Eq. 2 is supported by growth kinetics' data, SAXS and lattice calculations; the latter also providing a microscopic picture of the mechanism.




\begin{references}
\bibitem{8} Eq. 6 can be obtained from the full expression of $\sigma_e$, Eq. 19 in \cite{4}, considering that the fractional site occupation $\phi$ in the layer adjoining the folding plane is decreased by the fractional site occupation $\varphi$ of the  nucleant in such a way that $\phi+\varphi\approx$const., with $\varphi \ll \phi \approx 1$, without changing significantly the number of tight folds, Fig. 2 in \cite{4}, and finally considering that the segment chemical potential at constant $\phi$, Eq. 16 in \cite{4}, doesn't change significantly throughout the basal interphase.
\end{references}
\end{document}